\documentclass[12pt]{article}

\usepackage{amsmath}
\usepackage{amsfonts}
\usepackage{amssymb}
\usepackage{graphicx}
\usepackage{psfrag}
\usepackage{graphicx}
\usepackage{psfrag}

\newcommand{\be}{\begin{equation}}
\newcommand{\ee}{\end{equation}}
\newcommand{\ba}{\begin{eqnarray}}
\newcommand{\ea}{\end{eqnarray}}

\begin{document}
\begin{center}
{\bf
NEW IMPLICITLY SOLVABLE POTENTIAL PRODUCED BY SECOND ORDER SHAPE INVARIANCE}\\
\vspace{1cm}
{\large \bf F. Cannata$^{1,}$\footnote{E-mail: cannata@bo.infn.it},
M. V. Iof\/fe$^{2,}$\footnote{E-mail: m.ioffe@spbu.ru},
E. V. Kolevatova$^{2,}$\footnote{E-mail: e.v.krup@yandex.ru},\\
D. N. Nishnianidze}$^{3,2,}$\footnote{E-mail: cutaisi@yahoo.com}\\
\vspace{0.5cm}
$^1$ INFN, Via Irnerio 46, 40126 Bologna, Italy.\\
$^2$ Saint Petersburg State University,198504 Saint-Petersburg, Russia.\\
$^3$ Akaki Tsereteli State University, 4600 Kutaisi, Georgia.\\

\end{center}
\vspace{0.5cm}
\hspace*{0.5in}
\vspace{1cm}
\hspace*{0.5in}
\begin{minipage}{5.0in}
{\small
The procedure proposed recently by J.Bougie, A.Gangopadhyaya and J.V.Mallow \cite{mallow1} to study the general form of shape
invariant potentials in one-dimensional Supersymmetric Quantum Mechanics (SUSY QM) is generalized to the case
of Higher Order SUSY QM with supercharges of second order in momentum. A new shape invariant potential
is constructed by this method. It is singular at the origin, it grows at infinity, and its spectrum depends on the
choice of connection conditions in the singular point. The corresponding Schr\"odinger equation
is solved explicitly: the wave functions are constructed analytically, and the energy spectrum is defined implicitly
via the transcendental equation which involves Confluent Hypergeometric functions.}
\end{minipage}


\section*{\normalsize\bf 1. \quad Introduction.}
\vspace*{0.5cm}
\hspace*{3ex} Starting from the first paper of E.Witten \cite{witten} of 1981, the method of Supersymmetric Quantum Mechanics (SUSY QM) became a new effective tool \cite{cooper}, \cite{ai} for investigation of different problems of conventional Quantum Mechanics. From a mathematical point of view, this method - for one-dimensional space - is a reformulation of so called Darboux transformation of Sturm-Liuville equation  well known to mathematicians since more than hundred years \cite{darb}. SUSY QM approach was helpful for investigation of different problems in Quantum Mechanics and a lot of generalizations were proposed. In particular, the multi-dimensional generalization \cite{abe}, \cite{ioffe1}, \cite{ai} of SUSY QM which led to many new results, mostly for two-dimensional case \cite{david}, \cite{ioffe1} has to be mentioned. This multi-dimensional approach can be considered also as a generalization of old Darboux transformations.

Among new ideas provided by SUSY QM method, one - the shape invariance - is of special significance \cite{genden}, \cite{mallow1}, \cite{mallow}, \cite{shape2000}. It allows to solve some quantum models exactly or quasi-exactly. The list of known exactly solvable one-dimensional models is very restricted (approximately, about ten potentials are in this list). Supersymmetric transformations provide the partnership between pairs of potentials and therefore lead to new solvable potentials \cite{fernandez}, though sometimes of rather complicated analytical form. The method of shape invariance gives a very elegant {\it algebraic} algorithm to construct exactly solvable models without direct solution of differential equations. All previously known \cite{infeld} one-dimensional exactly solvable systems were shown to be shape invariant ones \cite{dabrowska}. Also, this approach provides analytical solution of several nontrivial two-dimensional problems \cite{new}, \cite{morse}.

Recently, the general investigation of additive form of shape invariance for superpotentials without explicit dependence on parameters was performed by J.Bougie et al \cite{mallow1}, \cite{mallow}. It was demonstrated that no additional shape invariant models in one-dimensional case can be constructed. This result was obtained in the framework of standard SUSY QM with supercharges of first order in derivatives. Meanwhile, it is known \cite{ais}, \cite{acdi-1}, \cite{higher} that such standard SUSY QM does not exhaust all
opportunities to fulfill the generalized SUSY algebra for Superhamiltonian $\hat H$ and supercharges $\hat Q^{\pm}$. Higher order supercharges are also possible, and for example supercharges of second order in momentum lead to some new results \cite{higher}.

In the present paper just a potential which fulfills the second order supersymmetry will be produced by the condition of shape invariance of second order. This potential is of very compact form, it depends on arbitrary parameter but has a strong singularity of $g/x^2$ kind with $g\in (-1/4, 0).$ This last property forces to construct a suitable self-adjoint extension of the Hamiltonian $H$, i.e. the suitable class of functions where $H$ acts. The paper is organized as follows. Section 2 contains a brief summary of one-dimensional SUSY QM and shape invariance. A shape invariant potential is built in Section 3 by means of second order supercharges. The direct analytical solution of the Schr\"odinger equation is given in Section 4. The problem of a suitable class of functions belonging to the domain of the self-adjoint extension of $H$ is discussed in Section 5 in terms of connection conditions at the origin. Section 6 includes derivation of the Spectrum Generating equation and its analysis, and finally, some conclusions are given in Section 7.

\section*{\normalsize\bf 2. \quad SUSY QM and Shape Invariance.}
\vspace*{0.5cm}
\hspace*{3ex} The main ingredients of one-dimensional SUSY Quantum Mechanics and of shape invariance will be briefly presented in this section (see details in Refs.\cite{cooper}).
An arbitrary one-dimensional Hermitian Schr\"odinger Hamiltonian with potential $V(x)$ and mass $m=1/2$ can be factorized:
\be
H = Q^+Q^- = -\hbar^2\partial^2 + V(x);\quad \partial\equiv \frac{d}{dx} \label{H}
\ee
by means of the first order differential operators $Q^{\pm};\quad Q^+=(Q^-)^{\dag}.$ Directly by construction, the superpartner Hamiltonian
\be
\tilde H = Q^-Q^+ = -\hbar^2\partial^2 + \tilde V(x) \label{tildeH}
\ee
is intertwined with initial $H:$
\be
H Q^+ = Q^+ \tilde H;\quad
Q^- H = \tilde H Q^-. \label{intertw}
\ee
Here, the first order supercharge operators $Q^{\pm}$ are expressed in terms of the superpotential function $W(x):$
\begin{equation}\label{WWW}
Q^{\pm} = \mp\hbar\partial + W(x),
\end{equation}
and the partner potentials are:
\be
V(x) = W^2(x)-\hbar W^{\prime}(x);\quad \tilde V(x) = W^2(x)+\hbar W^{\prime}(x). \label{V}
\ee
As a consequence of (\ref{intertw}), almost all wave functions of $H,\, \tilde H$ are interrelated:
\be
\Psi_n(x) = Q^+ \tilde\Psi_n(x);\quad
\tilde\Psi_n(x) = Q^- \Psi_n(x),\label{psi}
\ee
the only exclusions are possible normalizable zero modes of operators $Q^{\pm},$ but each of operators
admits not more than one such zero mode.

In some cases the partner Hamiltonians $H,\, \tilde H$ have more close relationship. Let us suppose \cite{genden}, \cite{mallow1}, \cite{mallow}, \cite{shape2000}, \cite{new} that both Hamiltonians (\ref{H}), potentials (\ref{V}),
superpotentials depend not only on coordinate $x,$ but also on some parameter $a,$ something like a coupling constant: $H=H(x; a);\, \tilde H=\tilde H(x; a),$ etc.
Let us suppose also that the potentials $V(x; a),\quad \tilde V(x, a)$ obey the so called shape invariance property:
\begin{equation}\label{shape}
V(x; a+\hbar) = \tilde V(x; a).
\end{equation}
Mathematically, the Planck constant $\hbar$ above can be considered above as a constant with arbitrary numerical value. Practically, in the Schr\"odinger equation context, the shape invariance of known solvable potentials is realized just by $\hbar$ shift of parameter (see details in \cite{mallow1}, \cite{mallow}). In essence,
the relation (\ref{shape}) means that
the dependence of superpartners $H(x; a),\, \tilde H(x; a)$ on coordinate $x$
is qualitatively close, but quantitatively slightly different. The distinction is determined by dependence of potentials $V,\, \tilde V$ on slightly different values
of parameter, i.e. "the shape" of potentials is the same. The potentials $V(x; a),$ which obeys the relation (\ref{shape}), are called shape invariant potentials. The list of such potentials is very short, and it practically coincides with the list of known solvable potentials. Usually, exact solvability of the Schr\"odinger equation with specific potential is established by exact solution of the corresponding Sturm-Liuville equation providing the analytic expressions for normalizable eigenfunctions and eigenvalues -- the wave functions and energies of quantum system. In contrast to this, shape invariance (\ref{shape}) together with intertwining relations (\ref{intertw}) provide different - purely algebraic - algorithm to find the spectrum and wave functions of the model.

The algebraic shape invariant algorithm gives the following prescription. If one chooses the ground state energy of $H(x, a)$ to be $E_0(a),$ the energy of $n-$th excited state of $H(x, a)$ is:
$$
E_n(a)= E_0(a+n\hbar)
,
$$
and its wave function is obtained from the ground state wave function as
\be
\Psi_n(x, a)=Q^+(x, a)Q^+(x, a+\hbar)...Q^+(x, a+(n-1)\hbar)\Psi_0(x, a+n\hbar) \label{psipsi}.
\ee
The ground state function $\Psi_0(x, a)$ is supposed to be known for arbitrary values of parameter $a.$ The shape invariance relation (\ref{shape}) and intertwining relations (\ref{intertw}) provide, for example:
\ba
&&H(x, a)Q^+(x,a)\Psi_0(x, a+\hbar) =
Q^+(x, a)\tilde H(x,a)\Psi_0(x, a+\hbar)=\nonumber\\
&=&
Q^+(x, a)H(x,a+\hbar)\Psi_0(x, a+\hbar)=E_0(a+\hbar)Q^+(x, a)\Psi_0(x, a+\hbar).   \nonumber
\ea
This chain of relations, obtained purely algebraically, shows that indeed, $\Psi_1(x, a)$ is an eigenfunction of $H(x, a)$ with energy $E_0(a+\hbar).$
Analogously, the expression (\ref{psipsi}) can be checked for arbitrary natural $n.$

In papers \cite{mallow} the most general dependence of the superpotential in (\ref{WWW}) on parameter $a$ was studied. It was proved that all conventional exactly solvable potentials, such as harmonic oscillator, Coulomb, Morse, P\"oschl-Teller, Scarf etc. can be reproduced by means of the shape invariance approach with $W=W(x, a).$ Meanwhile,
some recently constructed exactly solvable potentials \cite{quesne1} were missed in this algorithm. In paper \cite{mallow1}, a more general class of supercharges was considered, where the superpotential depends not only on parameter $a,$ but also explicitly on the constant $\hbar .$ Expanding the superpotential in powers of $\hbar ,$ potentials of \cite{quesne1} were also obtained as a particular case. In the next Section, the approach with this more general dependence of supercharges will be used, but in the context of higher order SUSY QM.

\section*{\normalsize\bf 3. \quad Construction of New Shape Invariant Potential.}
\vspace*{0.5cm}
\hspace*{3ex}The conventional SUSY QM admits the so called Higher Order generalization \cite{ais}, \cite{acdi-1}, \cite{higher} where the operators $Q^{\pm}$
are differential operators of order $N>1.$ It is known that arbitrary higher order transformation is equivalent to a series of first and second order elements.
In turn, the latter can be factorized onto two first order supercharges which intertwine the initial Hamiltonian with the intermediate Schr\"odinger-like Hamiltonian \cite{acdi-1}.
The intermediate system has real or complex potential depending on the sign of the constant parameter $d$ (see (\ref{Q}) - (\ref{b}) below), and in this sense second order supercharge is \cite{acdi-1} reducible or irreducible, correspondingly.

The case of supercharges of second order was studied completely \cite{acdi-1}, \cite{ai}, and just this case will be used in the present paper in the shape invariance context. The most general form of second order operators $Q^{\pm}$ is:
\begin{equation}\label{Q}
Q^+(x, a) = \hbar^2\partial^2 - 2\hbar f(x, a)\partial + b(x, a);\quad \partial\equiv d/dx
\end{equation}
It is known that the intertwining relations (\ref{intertw}) with supercharges (\ref{Q}) are fulfilled iff
\begin{equation}\label{VV}
V(x, a) = -2\hbar f^{\prime} + f^2 + \frac{\hbar^2}{2}(\frac{f^{\prime\prime}}{f}-\frac{(f^{\prime})^2}{2f^2})-\frac{d}{4f^2}+\gamma,
\end{equation}
where $f(x)$ is an arbitrary function of $x$ and function $b(x)$ is expressed in terms of $f(x):$
\be
b(x) = - \hbar f^{\prime} + f^2 - \frac{\hbar^2 f^{\prime\prime}}{2f} + \biggl(\hbar\frac{f^{\prime}}{2f}\biggr)^2 + \frac{d}{4f^2}.\label{b}
\ee
Both $d$ and $\gamma$ are arbitrary constants depending explicitly only on parameters $a$ and $\hbar ,$ in contrast to $f(x),$ which depends only on coordinate $x.$
This form of supercharges $Q^{\pm}$ is not the most general, but the choice of $a-$dependent functions $f(x)$ would lead to much more complicated problems.

Thus, the case with
\ba
f(x,a)=f(x),\qquad d(a,\hbar)=\sum_{n=0}^{\infty}\hbar^nd_n(a),\qquad \gamma(a,\hbar)=\sum_{n=0}^{\infty}\hbar^n\gamma_n(a) \nonumber
\ea
will be considered. Resummation of series gives:
\ba
&&d(a+\hbar,\hbar)-d(a,\hbar)=\sum_{n=1}^{\infty}\hbar^n\sum_{k=0}^{n-1}
\frac{d_k^{(n-k)}(a)}{(n-k)!}\equiv\sum_{n=1}^{\infty}\hbar^nD_n(a), \nonumber\\
&&\gamma(a+\hbar,\hbar)-\gamma(a,\hbar)=\sum_{n=1}^{\infty}\hbar^n\sum_{k=0}^{n-1}
\frac{\gamma_k^{(n-k)}(a)}{(n-k)!}\equiv\sum_{n=1}^{\infty}\hbar^nA_n(a). \nonumber
\ea
Then, the shape invariance condition (\ref{shape}) leads to relation:
\be
4\hbar f'(x)+\frac{\sum_{n=1}^{\infty}\hbar^nD_n(a)}{4f^2(x)}=\sum_{n=1}^{\infty}\hbar^nA_n(a). \nonumber
\ee
It is obvious now that:
\ba
&&4f'(x)+\frac{D_1(a)}{4f^2(x)}=A_1(a);         \label{99}\\
&&D_n(a)=A_n(a)=0, \qquad n>1.\label{100}
\ea
In turn, taking derivative of (\ref{99}) over parameter $a$ and separating variables, one obtains for nontrivial $f(x):$
\be
D_1(a)=const\equiv -\frac{16 c^3}{3},\quad A_1(a)=const\equiv \sigma \label{111}
\ee
where the definition of the constant for $D_1$ in the r.h.s. will be clear below (see Eq.(\ref{ff})). Eqs.(\ref{100}), (\ref{111}) give:
\be
d_n(a)=const\equiv d_n, \quad \gamma_n(a)=const\equiv \gamma_n, \quad n>1. \nonumber
\ee
By definition, $D_1(a)=d'_0(a),$ $A_1(a)=\gamma'_0(a),$ and the dependence has the form:
\be
d(a,\hbar)=-\frac{16 c^3}{3}a+\sum_{n=1}^{\infty}\hbar^nd_n\equiv -\frac{4c^3}{3}a+\kappa(\hbar),
\qquad \gamma(a,\hbar)=\sigma a + \sum_{n=1}^{\infty}\hbar^n\gamma_n\equiv \sigma a + \delta(\hbar),\label{133}
\ee
where the constants $c, \kappa, \sigma, \delta $ do not depend on the parameter $a.$

Thus, the potential (\ref{VV}) with constants $d,\,\gamma$ from (\ref{133}) is shape invariant if the function $f(x)$ satisfies equation:
\begin{equation}\label{f}
f^{\prime}(x)-\frac{c^3}{3f^2(x)}=\sigma/4 .
\end{equation}
The case of $c=0$ in (\ref{f}) corresponds to the well known potential - singular oscillator potential - and it will not be considered here. For all
nonzero values of constant $\sigma,$ equation (\ref{f}) can not be solved explicitly. This is a reason to choose the $\sigma =0,$ for which one obtains the monomial solution:
\begin{equation}\label{ff}
f(x)=cx^{1/3},
\end{equation}
and, after the convenient choice $\hbar\equiv 1$ (it can be fixed independently from the earlier choice $2m=1$), the superpartner potentials are as follows:
\begin{eqnarray}
  V(x) &=& \frac{2c(2a-1)}{3x^{2/3}}-\frac{5}{36x^2}+c^2x^{2/3}; \label{VVV} \\
\tilde V(x) &=& \frac{2c(2a+1)}{3x^{2/3}}-\frac{5}{36x^2}+c^2x^{2/3}, \nonumber
\end{eqnarray}
It is easy to check that by rescaling the coordinate $x\to c^{-3/4}x$ in both kinetic and potential terms of the Hamiltonian one can achieve $c\equiv 1.$ From now on this value of $c$ will be used, and thus the model depends only on one parameter $a.$ Potential (\ref{VVV}) is even function of $x,$ its form is shown on Figs. 1-3 for different values of parameter: $a=1,$ $a=2$ and $a=3.$

\vspace{10pt}

\begin{center}
\includegraphics[height=5cm]{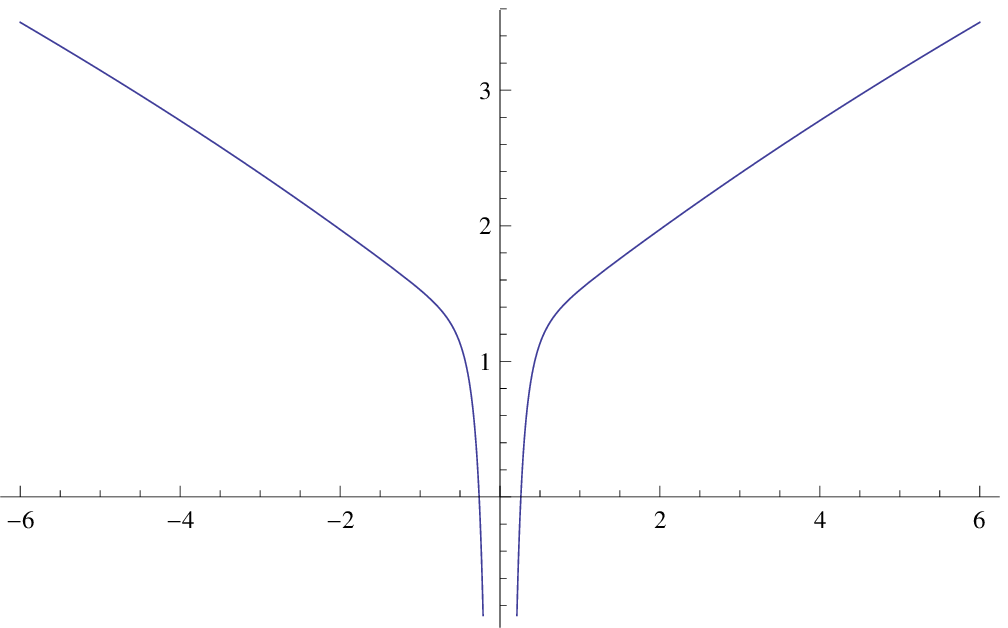}

\noindent{\it Fig.1} Plot of potential (\ref{VVV}) for $ a=1.$

\vspace{15pt}

\includegraphics[height=5cm]{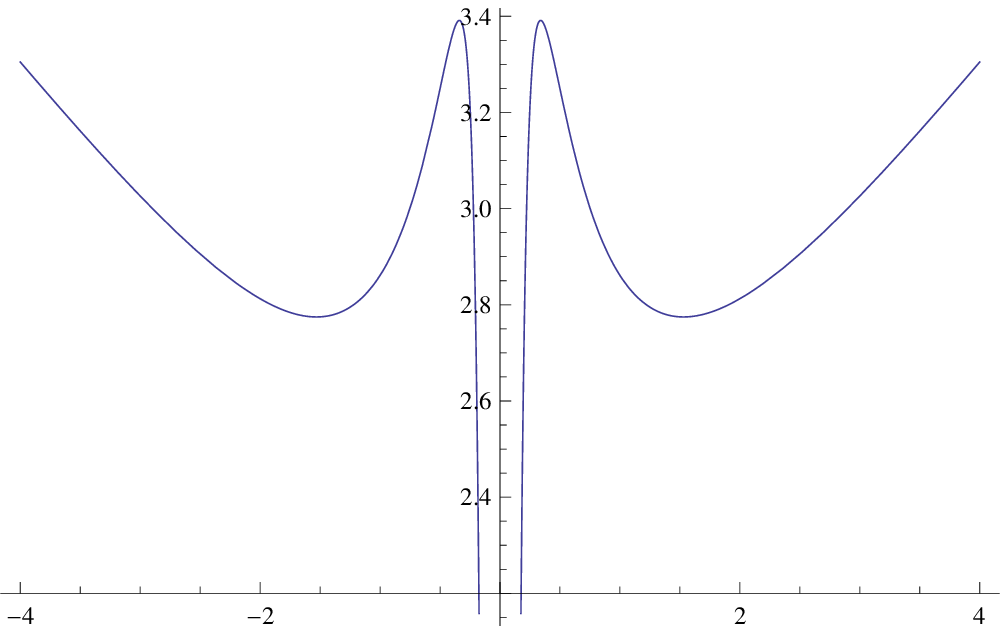}

\noindent{\it Fig.2} Plot of potential (\ref{VVV}) for $a=2.$

\vspace{15pt}
\includegraphics[height=5cm]{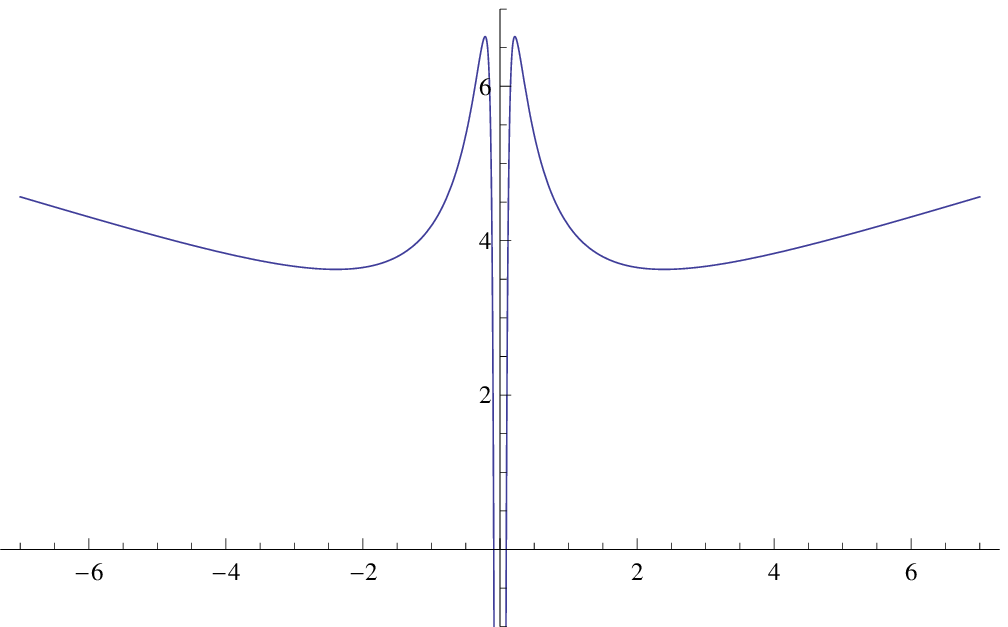}

\noindent{\it Fig.3} Plot of potential (\ref{VVV}) for $a=3.$

\end{center}

\vspace{10pt}

It is obvious that, the potential $V(x)$ is shape invariant in parameter $a,$ as it was expected by construction:
\begin{equation}\label{HH}
V(x; a+1)=\tilde V (x; a);\quad \gamma(a)=0.
\end{equation}
 As a remark, from the very beginning one could avoid the dependence of $d$ and $\gamma$ on $\hbar ,$ and to check directly that the dependence (\ref{133}) of these constants only on $a$ provides both the intertwining relations and shape invariance.

\section*{\normalsize\bf 4. \quad Direct Solution of the Schr\"odinger equation.}
\vspace*{0.5cm}
\hspace*{3ex} Our task is to investigate the potential $V(x)$ in (\ref{VVV}) for positive values of constant $a>0.$ This potential
has a strong but physically acceptable singularity $\sim 1/x^2$ at the origin: its coefficient is larger than $-1/4$ and prevents the "fall to the center" \cite{landau}. In this Section
we will demonstrate that the Schr\"odinger equation with this potential can be solved analytically in terms of confluent hypergeometric functions \cite{bateman}. Due to the singularity at $x=0,$ this solution
is not sufficient to provide the physical wave functions of the problem. Indeed, while the growing behaviour of potential (\ref{VVV}) at infinity corresponds to
the unique decreasing ($L_2-$integrable) solution, the situation at the origin is different. Namely, both independent solutions at the origin are decreasing, and this fact
does not allow to determine the discrete spectrum of the model. According to standard terminology \cite{reed}, the behaviour at $x\to\pm\infty$ is of the limit-point type, while the behaviour at $x\to 0$ is of the limit-circle one. How to deal with this problem will be studied in the next Section of the paper.

Let us look for the solution of the Schr\"odinger equation with potential (\ref{VVV}):
\be
H(x)\Psi(x)=E\Psi(x);\quad H(x)=-\partial^2 + V(x) \label{1}
\ee
in the following form:
\be
\Psi(x)=\Psi_0(x,a)\psi (x),\quad
\Psi_0(x,a)\equiv\exp\biggl(\int^x\frac{-2f^2(y)+f'(y)+\lambda(a)}{2f(y)}dy\biggr),   \label{3}
\ee
where according to (\ref{ff}) $f(x)=x^{1/3}.$ Choosing $\lambda (a)\equiv -\frac{4\sqrt{a}}{\sqrt{3}},$ the equation (\ref{1}) can be rewritten as:
\be
\biggl(\partial^2+\frac{f'(x)-2f^2(x)+\lambda(a)}{f(x)}\partial + E-\lambda(a)\biggr)\psi(x)=0. \label{4}
\ee
In the following, the variable
\be
\zeta=\sqrt{3}\bigl(f^2(x)-\frac{\lambda }{2}\bigr)   \label{5}
\ee
will be used instead of $x,$ and Eq.(\ref{4}) takes the form:
\be
\biggl(\frac{d^2}{d\zeta^2}-\zeta\frac{d}{d\zeta}+ \frac{\sqrt{3}}{4}(E-\lambda ) \zeta + \frac{3\lambda }{8}(E-\lambda )\biggr)\psi (\zeta)=0. \label{7}
\ee
Separating one more multiplier from $\psi (\zeta) :$
\be
\psi (\zeta)\equiv y(\zeta)\exp{\bigl(\frac{\sqrt{3}}{4}(E-\lambda )\zeta\bigr)},  \label{9}
\ee
one obtains new form of Eq.(\ref{7}):
\be
\biggl(\frac{d^2}{d\zeta^2}+\bigl(\frac{\sqrt{3}}{2}(E-\lambda )-\zeta\bigr)\frac{d}{d\zeta}+ \frac{3}{16}(E^2-\lambda^2 )\biggr)y(\zeta)=0.   \label{10}
\ee
Finally, the new transformation of variable:
\be
\xi=\frac{1}{2}\bigl(\zeta -\frac{\sqrt{3}}{2}(E-\lambda )\bigr)^2=\frac{3}{8}\bigl(2f^2(x)-E\bigr)^2.  \label{11}
\ee
reduces Eq.(\ref{10}) to the equation for the confluent hypergeometric function:
\be
\biggl(\xi\frac{d^2}{d\xi^2}+(\frac{1}{2}-\xi)\frac{d}{d\xi} - \alpha\biggr)y(\xi)=0,
\quad \alpha\equiv \frac{3}{32}(\lambda^2-E^2). \label{12}
\ee

It has two independent solutions:
\be
y_1(\xi)=F(\alpha,\frac{1}{2};\xi);\quad y_2(\xi)=\xi^{1/2}F(\alpha+\frac{1}{2},\frac{3}{2};\xi),  \label{13}
\ee
with $F$ - confluent hypergeometric functions \cite{bateman}. The general solution of initial equation (\ref{1}) is a linear combination ($N_{1, 2}-$arbitrary constants):
\be
\Psi (x,E)=N_1\Phi^{(1)}(x, E)+N_2\Phi^{(2)}(x, E), \label{1111}
\ee
with
\be
\Phi^{(i)}\equiv\Psi_0(x,a)\exp{(\frac{\sqrt{3}}{4}(E-\lambda )\zeta)}y_i(\xi(x))\sim \Psi_0(x,a)y_i(\xi(x))\exp{\biggl(\frac{3}{4}(E-\lambda )f^2(x)\biggr)}.\label{14}
\ee

The asymptotic behaviour of the confluent hypergeometric function for positive $\xi\to\infty$ is:
\be\label{F}
F(a,c;\xi)\sim \xi^{a-c} e^\xi \frac{\Gamma(c)}{\Gamma(a)},
\ee
where $\Gamma $ is the standard Gamma function. Thus, taking into account the $x\to\pm\infty$ behaviour of all multipliers in (\ref{14}), the asymptotics of $\Psi$ is:
\ba
\Psi (x,E) &\sim & N_1\frac{\Gamma(1/2)}{\Gamma(\alpha)}\exp{(\frac{3f^4}{2})}(\frac{3f^4}{2})^{\alpha -1/2} +
N_2\frac{\Gamma(3/2)}{\Gamma(\alpha +1/2)}(2f^2)\exp{(\frac{3f^4}{2})}(\frac{3f^4}{2})^{\alpha -1} \nonumber\\
&\sim &
\exp{(\frac{3f^4}{2})}(\frac{3f^4}{2})^{\alpha -1/2} \biggl[N_1\frac{\Gamma(1/2)}{\Gamma(\alpha)} + N_2\frac{2\sqrt{2}\Gamma(3/2)}{\sqrt{3}\Gamma(\alpha+1/2)}\biggr], \label{asymp}
\ea
where some constants were absorbed additionally in normalization constants $N_1, N_2.$
The condition of vanishing of the exponentially growing terms in asymptotics provides relation between constants $N_1, N_2:$
\be\label{NN}
\frac{N_1}{N_2} = -\frac{\Gamma(3/2)}{\Gamma(1/2)}\frac{\Gamma(\alpha)}{\Gamma(\alpha +1/2)}\frac{2\sqrt{2}}{\sqrt{3}}\equiv \gamma =
-\frac{\sqrt{2}\Gamma(\alpha)}{\sqrt{3}\Gamma(\alpha +1/2)},
\ee
and one of them is still arbitrary (up to normalization of wave function (\ref{1111})).

\section*{\normalsize\bf 5. \quad Connection conditions at the limit-circle singular point.}
\vspace*{0.5cm}
\hspace*{3ex} Thus, in the previous Section, the general $L_2-$integrable solution (\ref{1111}), (\ref{14}) of the Schr\"odinger equation for arbitrary value of energy $E$ was obtained up to an overall normalization constant. This knowledge does not provide the spectrum of the model: the energy $E$ does not appear to be constrained, while from the form of potential (\ref{VVV}) it is clear that the spectrum should be discrete, bounded from below and not bounded from above. The problem is well known since a long time \cite{zirilli} in the context of different physical models, and its origin is due to the properties of the Hamiltonian with potential (\ref{VVV}) around the singular point $x\to 0,$ which are reflected in the behaviour of solution (\ref{1111}) in this region.

The point $x=0,$ where both independent solutions are normalizable, is the so called limit-circle point. As for Hamiltonian of the model, it is not self-adjoint, and one has to find some self-adjoint extensions of $H,$ which are acceptable from physical and/or mathematical point of view. The task of such sort was considered in many papers, all of them describe the algorithms to find the self-adjoint extensions. It was proved that the entire variety of extensions forms the two-parameter set described by elements of unitary group $U(2).$ We will follow one of such algorithms - due to papers \cite{tsutsui-main}  - formulated in terms of connection conditions in the limit-circle singular point (see also the preceding papers \cite{fulton}.

Below the scheme of construction of the entire variety of self-adjoint extensions of given Hamiltonian with singularity of limit-circle type at the origin will be briefly presented. The first step of the procedure is to construct a pair $\varphi^{(i)}(x)$ of independent auxiliary solutions (reference states) of the Schr\"odinger equation (\ref{1}) with some (arbitrary) real energy value $\epsilon .$  The normalizability of reference states at infinity is not assumed, and their behaviour at the origin is determined by the singularity of potential - in our case of potential (\ref{VVV}) independent solutions have the following behaviour at the origin: $\sim x^{1/6}$  and $\sim x^{5/6}.$

Let us suppose that the auxiliary solutions $\varphi^{(i)}(x)$ of (\ref{1}) are known, and that their Wronskian is normalized:
\be
W[\varphi^{(1)}, \varphi^{(2)}]\equiv \varphi^{(1) \prime}\varphi^{(2)}-\varphi^{(1)}\varphi^{(2) \prime} = 1. \nonumber
\ee
Let $\Psi(x)$ and $\tilde\Psi(x)$ to be arbitrary decreasing at infinity functions for which differential operator is well defined.
The condition of symmetric property of $H$ reads:
\be\label{sym}
\int_X dx\bigl[\tilde\Psi^{\star}(x)H(x)\Psi(x)\bigr] = \int_X dx\bigl[(H(x)\tilde\Psi(x))^{\star}\Psi(x)\bigr].
\ee
Integrating (\ref{sym}) by parts and taking into account that both $\Psi$ and $\tilde\Psi$ are exponentially decreasing at infinity, one obtains "the boundary terms" only from the left and right vicinities of $x=0:$
\be\label{WW}
0 = \int_X dx\bigl[\tilde\Psi^{\star}(x)H(x)\Psi(x)\bigr] - \int_X dx\bigl[(H(x)\tilde\Psi(x))^{\star}\Psi(x)\bigr]=\biggl( W[\tilde\Psi^{\star}, \Psi ]_{+0}
-W[\tilde\Psi^{\star}, \Psi ]_{-0}\biggr).
\ee
The r.h.s. of (\ref{WW}) can be expressed using the auxiliary solutions $\varphi^{(i)}$ via the chain of the following transformations of Wronskians:
\ba
W[\tilde\Psi^{\star}, \Psi ]=\left|
                               \begin{array}{cc}
                                 \tilde\Psi^{\star} & \tilde\Psi^{\star\prime} \\
                                 \Psi & \Psi^{\prime} \\
                               \end{array}
                             \right| =
                             \left|
                               \begin{array}{cc}
                                 \tilde\Psi^{\star} & \tilde\Psi^{\star\prime} \\
                                 \Psi & \Psi^{\prime} \\
                               \end{array}
                             \right|
                             \left|
                               \begin{array}{cc}
                                 \varphi^{(1) \prime} & \varphi^{(2) \prime} \\
                                 -\varphi^{(1)} & -\varphi^{(2)} \\
                               \end{array}
                             \right| = \nonumber\\
                            = \left|
                               \begin{array}{cc}
                                 \tilde\Psi^{\star} \varphi^{(1) \prime} - \tilde\Psi^{\star\prime} \varphi^{(1)} & \tilde\Psi^{\star} \varphi^{(2) \prime} - \tilde\Psi^{\star\prime} \varphi^{(2)} \\
                                 \Psi \varphi^{(1) \prime} - \Psi^{\prime} \varphi^{(1)} & \Psi \varphi^{(2) \prime} - \Psi^{\prime} \varphi^{(2)}  \\
                               \end{array}
                             \right| =
                                                          \nonumber\\
                             = W[\tilde\Psi^{\star}, \varphi^{(1)} ] W[\Psi , \varphi^{(2)}] -
                             W[\tilde\Psi^{\star}, \varphi^{(2)} ] W[\Psi , \varphi^{(1)}]. \nonumber
\ea
This transformation allows to rewrite the r.h.s. of (\ref{WW}) as:
\be\label{Omega}
\tilde\Omega^{(1) \dag} \Omega^{(2)} -\tilde\Omega^{(2) \dag} \Omega^{(1)}
\ee
where the two-component columns are defined as:
\be
\Omega^{(1)}\equiv \left(
                     \begin{array}{c}
                       W[\Psi, \varphi^{(1)}]_{+0} \\
                        W[\Psi, \varphi^{(1)}]_{-0} \\
                     \end{array}
                   \right); \quad \Omega^{(2)}\equiv \left(
                     \begin{array}{c}
                       W[\Psi, \varphi^{(2)}]_{+0} \\
                        - W[\Psi, \varphi^{(2)}]_{-0} \\
                     \end{array}
                   \right), \nonumber
\ee
and analogously, $\tilde\Omega^{(i)}$ are defined by replacing $\Psi \to \tilde\Psi .$ For $\Psi$ and $\tilde\Psi$ in a self-adjoint domain the expression (\ref{Omega}) has to be zero. This condition can be rewritten in terms of
\be
\Omega^{(\pm)}\equiv \Omega^{(1)}\pm i\omega\Omega^{(2)};\quad \tilde\Omega^{(\pm)}\equiv \tilde\Omega^{(1)}\pm i\omega\tilde\Omega^{(2)}, \label{omega}
\ee
with an arbitrary nonzero constant $\omega .$ For coinciding $\Psi\equiv\tilde\Psi ,$ the vanishing of (\ref{Omega}) is equivalent to the relation:
\be\label{U}
U\Omega^{(+)}=\Omega^{(-)}; \quad U\in U(2).
\ee
For the general case of different $\Psi$ and $\tilde\Psi ,$ but with the same matrix $U,$ the difference (\ref{Omega}) actually vanishes, and thus the symmetric self-adjoint operator $H$ is completely characterized by matrix $U,$ which may be therefore called characteristic matrix of the self-adjoint Hamiltonian.

The scheme described above can be directly applied to our model with potential (\ref{VVV}), starting from the convenient choice of the reference modes $\varphi^{(i)}, i=1,2 :$
\ba
&&\varphi^{(1)}\equiv -\frac{3}{4}\Phi^{(1)}(x, \epsilon\equiv\lambda)=-\frac{3}{4}\Psi_0(x, \lambda)\quad for\,\, x>0; \quad\varphi^{(1)}(-x)\equiv -\varphi^{(1)}(x); \label{mode1}\\
&&\varphi^{(2)}\equiv  \exp{(-4a)}\Phi^{(2)}(x, \epsilon\equiv\lambda)=\exp{(-4a)}(2f^2(x)-\lambda)\Psi_0(x, \lambda)F(1/2,3/2;\xi);\nonumber\\
&&\varphi^{(2)}(-x)\equiv  \varphi^{(2)}(x), \label{mode2}
\ea
where constant multipliers were chosen so to provide the value of Wronskian: $W[\varphi^{(1)}, \varphi^{(2)}] = 1.$ It will be clear below that the freedom in this choice does not affect the final results. Indeed, for diagonal $U$ equations for spectrum of the next Section does not depend on these constants, and for nondiagonal $U$ the change of constants can be compensated by change of $\omega $ in (\ref{omega}).

The next step is to choose the specific matrix $U\in U(2),$ which defines the symmetric operator $H.$ For example, the class of operators with diagonal $U$ can be selected. Namely, physically it corresponds to the vanishing probability flow through the point $x=0,$ since for such $U$, the boundary conditions at $x\to +0$ are not mixed with those for $x\to -0:$
\ba
j(x=0)&=&\frac{\hbar}{2im}(\Psi^*(x)\Psi^{\prime}(x)-\Psi^{*\prime}(x)\Psi(x))=\nonumber\\
&=&\frac{\hbar}{2im}\biggl(W[\Psi^*,\phi^{(1)}]W[\Psi ,\phi^{(2)}] - W[\Psi^*,\phi^{(2)}]W[\Psi ,\phi^{(1)}] \biggr)=0. \nonumber
\ea

Below the simplest ansatzes $U=\pm I$ will be considered. For the choice $U=-I,$ the first two-component column has to vanish $\Omega^{(1)} =0.$ Since both solutions $\Phi^{(i)}(x,E)$ are even functions of $x$ and $\varphi^{(i)}$ are also of definite parity, the components in $\Omega^{(1)}$ differ only by sign from each other. Thus, the Hamiltonian $H$ is self-adjoint in the domain of functions $\Psi$ satisfying the only - on the right side - condition:
\be\label{equation}
W[\Psi(x,E), \varphi^{(1)}(x)]_{+0}=0,
\ee
which realizes the connection condition for $\Psi(x, E)$ in the limit-circle singular point. This equation depends on energy $E,$ and it will determine the allowed values
$E_n,$ i.e. the discrete spectrum of the model. The wave functions (\ref{1111}), specified by (\ref{14}) and reference mode (\ref{mode1}) must be substituted into (\ref{equation}):
\be\label{eq}
N_1\cdot W\bigl[\Phi^{(1)}(x,E_n), \varphi^{(1)}(x)\bigr]_{+0}+N_2\cdot W\bigl[\Phi^{(2)}(x,E_n), \varphi^{(1)}(x)\bigr]_{+0}=0.
\ee
Due to $x\to\pm\infty$ asymptotic, one of the constants $N_1, N_2$ is expressed in terms of second (see (\ref{NN})), which is still arbitrary up to normalization.

\section*{\normalsize\bf 6. \quad Derivation of the Spectrum Generating Equation.}
\vspace*{0.5cm}
\hspace*{3ex} Practically, to find the Spectrum Generating Equation (SGE) for $U=-I$ one has to rewrite Eq.(\ref{eq}) by substituting functions $\Phi^{(i)}$ and $\varphi^{(1)}$ from (\ref{14}) and (\ref{mode1}). The expression $\varphi^{(1)}$ allows to obtain the $x\to +0$ limit of Wronskians in (\ref{eq}) explicitly as linear combinations of Confluent Hypergeometric Functions:
\ba
&&W[\Phi^{(1)}(x, E), \varphi^{(1)}(x)]_{+0} = -\frac{3}{8}\biggl[(E-\lambda) F(\alpha, 1/2; \frac{3E^2}{8})-\nonumber\\
&&-4\alpha E F(\alpha +1, 3/2; 3E^2/8) \biggr];
\nonumber\\
&&W[\Phi^{(2)}(x, E), \varphi^{(1)}(x)]_{+0} = -\frac{3E(E-\lambda )}{8} F(\alpha +1/2, 3/2; \frac{3E^2}{8})-\nonumber\\
&&-F(\alpha +1/2, 3/2; \frac{3E^2}{8}) -
\frac{3E^2(2\alpha +1)}{8} F(\alpha +3/2, 5/2; \frac{3E^2}{8}). \nonumber
\ea
Both Wronskians can be essentially simplified by means of recurrency relations between Confluent Hypergeometric Functions:
\ba
&&2\alpha F(\alpha +1,3/2;\frac{3E^2}{8}) = (2\alpha -1) F(\alpha ,3/2;\frac{3E^2}{8}) + F(\alpha ,1/2;\frac{3E^2}{8});
\nonumber \\
&& E^2 F(\alpha +3/2,5/2;\frac{3E^2}{8}) = 4\biggl( F(\alpha +3/2 ,3/2;\frac{3E^2}{8}) - F(\alpha +1/2 ,3/2; \frac{3E^2}{8})\biggr), \nonumber
\ea
and the following expressions are obtained:
\ba
&&W[\Phi^{(1)}(x, E), \varphi^{(1)}(x)]_{+0} = \frac{3}{4} \biggl[\frac{E+\lambda }{2} F(\alpha, 1/2; \frac{3E^2}{8}) + (2\alpha -1) E F(\alpha , 3/2; \frac{3E^2}{8})  \biggr]; \nonumber\\
&&W[\Phi^{(2)}(x, E), \varphi^{(1)}(x)]_{+0} = \frac{3E(E-\lambda )}{8} F(\alpha +1/2, 3/2; \frac{3E^2}{8})- F(\alpha +1/2, 1/2; \frac{3E^2}{8}). \nonumber
\ea
After that, taking into account (\ref{NN}), the SGE (\ref{eq}) for $U=-I,$  takes the form:
\ba
&&\Gamma(\alpha)\biggl[(2\alpha -1) y F(\alpha, 3/2; y^2)+\frac{1}{2}(y+\eta )F(\alpha, 1/2; y^2)\biggr] -\nonumber\\
&&-\Gamma(\alpha +1/2)\biggl[y (y-\eta ) F(\alpha +1/2, 3/2; y^2) - F(\alpha +1/2, 1/2; y^2)\biggr]
= 0,\label{eeq}
\ea
where new variables are used:
\be
y\equiv\frac{\sqrt{3} E}{2\sqrt{2}};\quad \eta\equiv\frac{\sqrt{3} \lambda}{2\sqrt{2}}, \label{def}
\ee
and the following constraint must hold:
\be\label{constr}
4\alpha=(\eta^2-y^2).
\ee

The obtained equation (\ref{eeq}), from which one has to determine solutions $y_n,$
and therefore, by (\ref{def}) - the allowed values of energy $E_n,$ is transcendental equation. In contrast to well-known exactly solvable models,
these values can not be written as an explicit function of $n.$ In this sense the model is similar to the standard quantum problem of a particle in the
rectangular well of finite depth, which is considered traditionally in many textbooks. The only opportunity to solve such sort of problems is to use some appropriate numerical
methods.

In the present case, the suitable tool is to explore the computer algebra system MATHEMATICA \cite{wolfram}. To simplify this task, it is technically useful to divide both sides of (\ref{eeq}) by a product $\Gamma (\alpha) \Gamma (\alpha + 1/2),$ this is equivalent transformation for $\alpha \neq -k/2;\,\, k=0, 1, 2, ... ,$ where this product is infinite. One can fix some values for parameters to illustrate the solution of SGE. The choice $\eta = -1$ will not be considered here since it corresponds to parameter $a=1/2,$ leading to potential (\ref{VVV}) without first term. Choosing, for example, $\eta = -2$ or $\eta = -3,$ values of several lowest $y_n$ can be determined with good accuracy from the plot of l.h.s. of (\ref{eeq}) after mentioned division by Gamma's. The result is given at Figs.4-5, where the discrete values $y_n$ (proportional to energy eigenvalues $E_n$) are plotted as functions of the level number $n$ for $\eta =-2$ and $\eta =-3.$

\begin{center}
\includegraphics[height=5cm]{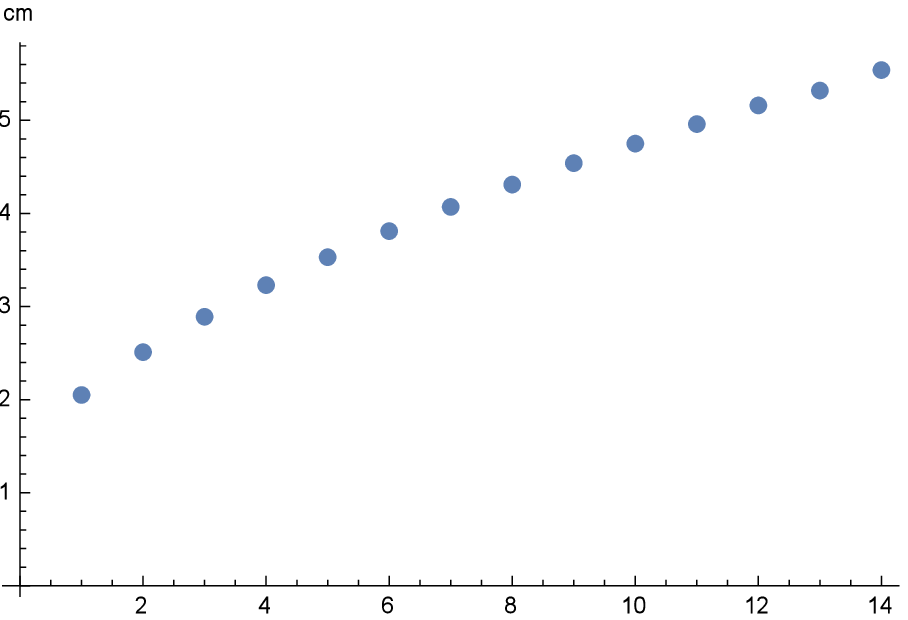}

\noindent{\it Fig.4} Discrete values $y_n$ as function of $n$ for $\eta =-2.$

\vspace{10pt}
\includegraphics[height=5cm]{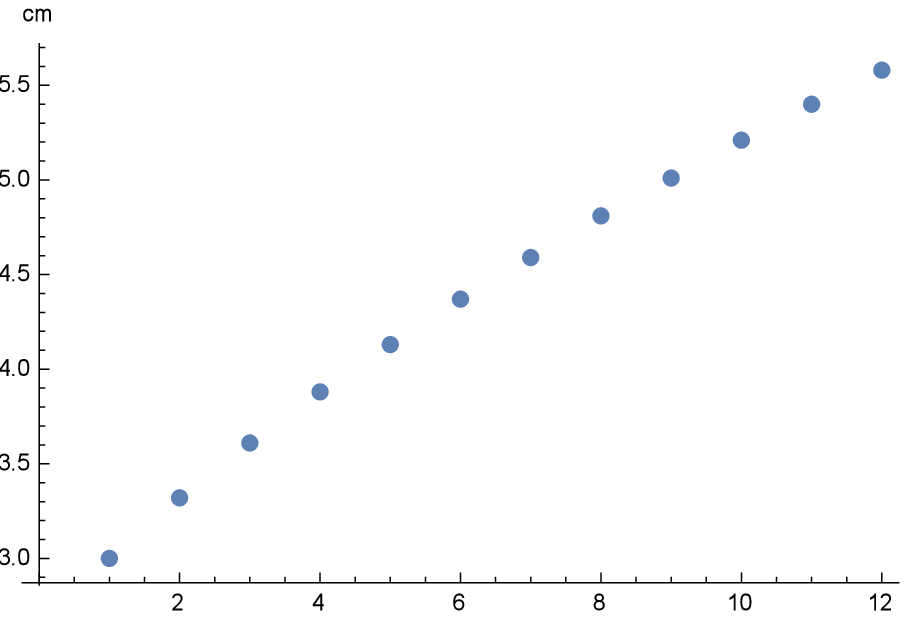}

\noindent{\it Fig.5} Discrete values $y_n$ as function of $n$ for $\eta =-3.$
\vspace{10pt}
\end{center}
One can notice that the first level $n=1$ on both plots corresponds to $y_1=\eta,$ i.e. according to (\ref{def}), to $E_1=\lambda .$ This observation is universal, since one can
derive from the expression (\ref{14}) that $\Phi^{(1)}(x)\sim\Psi_0(x, a)$ for $E=\lambda $ and $\alpha =0,$ and it has no nodes for finite $x.$ For $\alpha =0,$ relation (\ref{NN})
is fulfilled only if the constant $N_2=0.$ Comparing this argumentation with SGE, one can conclude that the second term in Eq.(\ref{eq}) vanishes, and therefore, Eq.(\ref{eeq}) contains only first term. For $\alpha =0$ and $y=\eta$ this term is zero, and SGE is fulfilled. Thus the states with $E_1=\lambda$ are the physical ground states in the case $U=-I.$
Concerning first excited states, one can check by means of MATHEMATICA that for $\eta =-2$ their wave functions with $E_2, E_3, E_4$ from Fig.4 have just $1, 2, 3$ nodes for finite $x.$ All these levels for $\eta = -2,$ starting from the ground state $E_1$ can be seen at the Fig.6, where the l.h.s. of SGE (\ref{eeq}) divided by product of Gamma functions is plotted. The positions of roots coincide with that on Fig.4.

\begin{center}

\includegraphics[height=5cm]{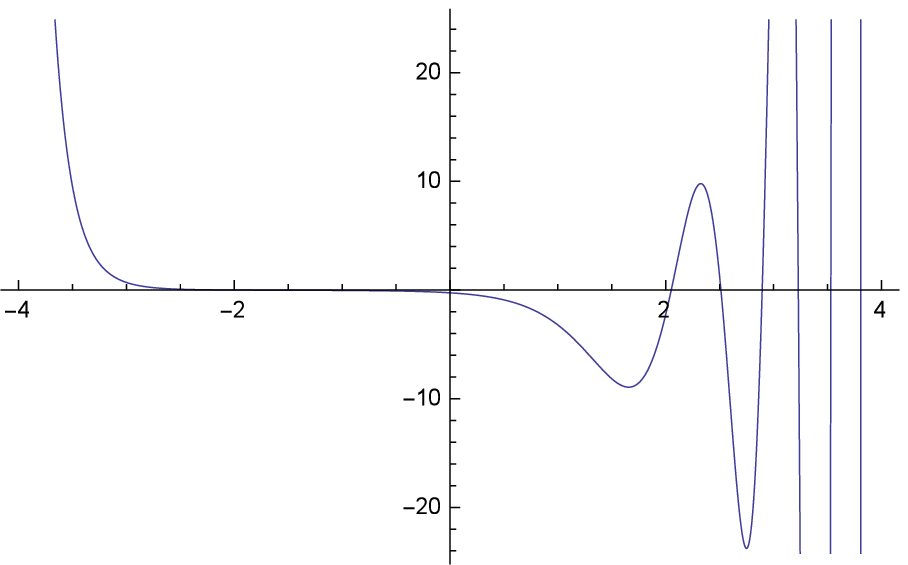}

\noindent{\it Fig.6} The plot of l.h.s. of (\ref{eeq}) for $\eta =-2$ divided by $\Gamma (\alpha) \Gamma (\alpha +1/2).$ Nodes correspond to the discrete spectrum $y_n.$

\vspace{10pt}
\end{center}

The analogous procedure can be performed for the case $U=+I.$ In this case, Wronskians with $\varphi^{(2)},$ instead of $\varphi^{(1)},$ must be calculated. The final result for the SGE is again the transcendental equation:
\ba
&&\frac{1}{\Gamma(\alpha +1/2)}\biggl\{[\eta(\eta +y)F(\alpha,1/2;y^2)+
2y\eta(2\alpha -1)F(\alpha,3/2;y^2)]F(1,3/2;-\eta^2)-\nonumber\\
&&-F(\alpha,1/2;y^2)\biggr\}-\nonumber\\
&&-\frac{2}{\Gamma(\alpha)}\biggl\{[\eta y(y-\eta)F(\alpha +1/2,3/2;y^2)-
\eta F(\alpha +1/2,1/2;y^2)]F(1,3/2;-\eta^2)+\nonumber\\
&&+yF(\alpha +1/2,3/2;y^2)\biggr\}=0. \label{sge-2}
\ea

\section*{\normalsize\bf 7. \quad Conclusions.}
\vspace*{0.5cm}
\hspace*{3ex} According to \cite{ais}, \cite{acdi-1}, an arbitrary supercharge operator (\ref{Q}) allows factorization onto two first order supercharges with real or complex intermediate potential depending on the sign of constant $d.$ For the pair of superpartner potentials (\ref{VVV}), factorization of second order supercharges (\ref{Q}) as in \cite{ais}, \cite{acdi-1} gives the intermediate potential:
\be
V_{Int}(x )= f^2-\frac{f''}{2f}+\frac{3}{4}\frac{f'^2}{f^2}+\frac{\lambda^2}{4f^2}+\frac{\lambda f'}{f^2}=x^{2/3}+\frac{7}{36x^2}+\frac{\lambda}{x^{4/3}}+\frac{4a}{3x^{2/3}},
\label{interm}
\ee
which is real. It is obvious that no shape invariance between the systems with potentials (\ref{VVV}) and (\ref{interm}) is observed. In this sense, the shape invariance between systems with potentials (\ref{VVV}) is irreducible.
The intermediate Hamiltonian with potential (\ref{interm}) is intertwined with the initial $H$ by means of first order operators of the form (\ref{WWW}) with superpotential \cite{acdi-1}:
\be
W(x, a) = f(x)-\frac{f^{\prime}+\sqrt{-d(a)}}{2f(x)}
\label{a}
\ee
Necessary to mention here, that besides of \cite{sukhatme}, where mainly the reflectionless potentials with two-step shape invariance were considered, the shape invariance of second order was studied also in a few recent papers \cite{su}, \cite{tanaka}. The first of them deal with factorized supercharges and the superpotential $W$ linearly depending on parameter $a$ of shape invariance. In contrast to this approach, in the present paper in terms of $W$ we considered actually the "square root dependence" of $W$ via $\sqrt{-d}\equiv\sqrt{a}$ (see (\ref{133})). As for the \cite{tanaka}, many potentials with two-step shape invariance were obtained there but most of them can not be written explicitly - only in terms of rather complicate functions $z(x).$

The main results of the present paper are the following. Using the shape invariance approach in the framework of second order SUSY QM, a new one-dimensional shape invariant model was built, and its shape invariance is irreducible. The corresponding Schr\"odinger equation admits directly the exact solution. Due to the singular properties of the obtained potential, the model is well defined only after the suitable choice of the connection conditions at the singular point. These conditions were rewritten explicitly as specific Spectrum Generating Equations (SGE) for two simplest variants $U=\pm I.$ The SGE equations are transcendental equations for the discrete spectrum ${E_n}$, which can be solved only numerically. Although MATHEMATICA is not able to calculate energies of very high excited states, the spectrum of the problem is evidently infinite in positive direction. The analysis of different terms in (\ref{eeq}) in the context of their properties under reflection $y\to -y$ shows that this equation having unbounded positive spectrum of $y_n$ can not have simultaneously very deep negative energies in the spectrum, confirming no fall to the center.

\section*{\bf Acknowledgments.}

The work was partially supported by the grants of Saint Petersburg State University N 11.38.660.2013, N 11.42.1303.2014 (M.V.I.) and by the grant RFBR N 13-01-00136-a (M.V.I. and E.V.K.). The work of D.N.N. was partially supported by the grant ATSU N 31. M.V.I. and D.N.N. acknowledge hospitality by INFN and University of Bologna where part of this work was done.

\end{document}